\documentclass[letter]{aa}  
\usepackage{graphicx}
\usepackage{float}

\usepackage{txfonts}
\newcommand{\OH}{OH~119~$\mu$m}
\newcommand{\nad}{\ion{Na}{I}~D}
\newcommand{\cii}{[\ion{C}{II}]}
\newcommand{\ci}{[\ion{C}{I}]}
\newcommand{\qsoname}{ULAS~J131911+095051}

\begin{document} 

   \title{Molecular Gas Inflows and Outflows in Ultraluminous Infrared Galaxies at $z\sim0.2$ and one QSO at $z=6.1$\thanks{{\it Herschel} is an ESA space observatory with science instruments provided by European-led Principal Investigator consortia and with important participation from NASA.}}
   \titlerunning{OH Molecular gas inflows and outflows}

   \author{R. Herrera-Camus\inst{1},
          E. Sturm\inst{2}, 
          J. Graci\'a-Carpio\inst{2},
          S. Veilleux\inst{3,4,5,6},
          T. Shimizu\inst{2},
          D. Lutz\inst{2},
          M. Stone\inst{3},
          E. Gonz\'alez-Alfonso\inst{7},
          R. Davies\inst{2},
          J. Fischer\inst{8},
          R. Genzel\inst{2},
          R. Maiolino\inst{5,6},
          A. Sternberg\inst{9},
          L. Tacconi\inst{2},
          \and
          A. Verma\inst{10}
	 }
	\authorrunning{Herrera-Camus et al.}
	
   \institute{Departamento de Astronom\'ia, Universidad de Concepci\'on, Barrio Universitario, Concepci\'on, Chile\\
              \email{rhc@astro-udec.cl}
              \label{1}
         \and
	Max-Planck-Institut f\"ur extraterrestrische Physik, Giessenbachstr., 85748 Garching, Germany
	\label{2}
\and
	Department of Astronomy and Joint Space-Science Institute, Univ. of Maryland, 
	College Park, MD 20742, USA 
	\label{3}
\and
	Space Telescope Science Institute, Baltimore, MD 21218
	\label{4}
\and
	Institute of Astronomy, University of Cambridge, Madingley Road, Cambridge CB3 0HA, UK
	\label{5}
\and
	Kavli Institute for Cosmology, University of Cambridge, Madingley Road, Cambridge CB3 0HA, UK
	\label{6}
\and
	Departamento de F\'isica y Matem\'aticas, Univ. de Alcal\'a, Campus Universitario, E-28871 Alcal\'a de Henares, Madrid, Spain
	\label{7}
\and
	George Mason University, Department of Physics \& Astronomy, MS 3F3, 4400 University Drive, Fairfax, VA 22030, USA
	\label{8}
\and
	Raymond and Beverly Sackler School of Physics \& Astronomy, Tel Aviv University, Ramat Aviv 69978, Israel
	\label{9}
\and
	Sub-department of Astrophysics, University of Oxford, Keble Road, Oxford, OX1 3RH
	\label{10}
}
 
  \abstract
  % context heading (optional)
  % {} leave it empty if necessary  
  {}
  % aims heading (mandatory)
  {We aim to search and characterize inflows and outflows of molecular gas in four ultraluminous infrared galaxies (ULIRGs; $L_{\rm IR}>10^{12}$~L$_{\odot}$) at $z\sim0.2-0.3$ and one distant quasi-stellar object (QSO) at $z=6.13$. 
  }
  % methods heading (mandatory)
   {We use {\it Herschel} PACS and ALMA Band~7 observations of the hydroxyl molecule (OH) line at rest-frame wavelength 119~$\mu$m which in absorption can provide unambiguous evidence for inflows or outflows of molecular gas in nuclear regions of galaxies. Our study contributes to double the number of \OH\ observations of luminous systems at $z\sim0.2-0.3$, and push the search for molecular outflows based on the \OH\ transition to $z\sim6$.
   }
  % results heading (mandatory)
   {We detect \OH\ high-velocity absorption wings in three of the four ULIRGs. In two cases, IRAS~F20036-1547 and IRAS~F13352+6402, the blueshifted absorption profiles indicate the presence of powerful and fast ($\sim200-500$~km~s$^{-1}$) molecular gas outflows. Consistent with an inside-out quenching scenario, these outflows are depleting the central reservoir of star-forming molecular gas at a similar rate than the intense star formation activity. In the case of the starburst-dominated system IRAS~10091+4704, we detect an inverted P-Cygni profile that is unique among ULIRGs and indicates the presence of a fast ($\sim400$~km~s$^{-1}$) inflow of molecular gas at a rate of $\sim100~M_{\odot}~{\rm yr}^{-1}$ towards the central region. Finally, we tentatively detect ($\sim3\sigma$) the \OH\ doublet in absorption in the $z=6.13$ QSO \qsoname. The \OH\ feature is blueshifted with a median velocity that suggests the presence of a molecular outflow, although characterized by a modest molecular mass loss rate of $\sim200~M_{\odot}~{\rm yr}^{-1}$. This value is comparable to the small mass outflow rates found in the stacking of the \cii\ spectra of other $z\sim6$ QSOs and suggests that ejective feedback in this phase of the evolution of \qsoname\ has subsided.
   }
   {}


 \maketitle
%
%________________________________________________________________

\section{Introduction}

Galaxies form and evolve driven by the complex exchange of regular matter with their surrounding medium in what is commonly known as the baryon cycle. 
One key element in this process is the flow in and out of galaxies of molecular gas, the fuel for star formation. On one hand, accretion of gas is required to sustain the observed star formation rate (SFR) in galaxies across cosmic time \citep[e.g.,][]{rhc_tacconi10,rhc_daddi10,rhc_bouche10,rhc_lagos15,rhc_genzel15,rhc_tacconi18}. On the other hand, strong outflows have the potential to remove a large fraction of the molecular gas reservoir which may result in the suppression of the star formation activity \citep[e.g.,][]{rhc_feruglio10,rhc_cicone14,rhc_veilleux17,rhc_fiore17,rhc_rhc19,rhc_fluetsch19,rhc_lutz19,rhc_rhc19}.

Finding and characterizing galactic molecular outflows in local and distant galaxies remains a challenge. One recent method that has proved to be successful is to search for P~Cygni profiles or high-velocity absorption wings in OH infrared spectra. {\it Herschel}, and more recently ALMA, have enabled systematic studies of molecular gas inflows and outflows traced by OH in nearby galaxies and (U)LIRGs \citep[e.g.,][]{rhc_fischer10,rhc_sturm11,rhc_spoon13,rhc_veilleux13,rhc_gonzalez-alfonso14,rhc_stone16,rhc_gonzalez-alfonso17}, and high redshift starbursts \citep[e.g.,][]{rhc_george14,rhc_zhang18,rhc_spilker18}. These studies find powerful molecular outflows with velocities that range from a few hundred to $\sim1000$~km~s$^{-1}$ and mass loss rates that are comparable or in some cases up to $\sim10\times$ higher than the star formation rates. 

\begin{table*}[t!]
\begin{center}
\caption{Properties of the galaxies in the sample\label{tab:sample}}
\begin{tabular}{cccccccccc}
\hline \hline
Source & $z$ & WISE & $L_{\rm FIR}$ & $v_{50}$ & $v_{84}$ & EW$_{abs}$ & EW$^{v<-200}_{abs}$ & Flux$_{abs}$ & $\dot{M}_{\rm mol}$ \\
 & & Class. & ($10^{12}~L_{\odot}$) & (km~s$^{-1}$) & (km~s$^{-1}$) &  (km~s$^{-1}$) & (km~s$^{-1}$) & (Jy~km~s$^{-1}$) & ($M_{\odot}~{\rm yr}^{-1}$) \\
\hline
IRAS~F13352+6402 & 0.2360\tablefootmark{a} & Sy/QSO& 2.9\tablefootmark{d}  & $-115$ & $-480$ & 338 & 115 & 182$\pm$9 & $\sim570$\tablefootmark{g} \\
IRAS~F10091+4704 & 0.2451\tablefootmark{b} & SB & 3.7\tablefootmark{d} & $+430$ & $+75$ & 211 & $-$ & 112$\pm$7 & $\sim100$\tablefootmark{h}\\
IRAS~F20036-1547 & 0.1920\tablefootmark{a} & Sy/QSO & 2.9\tablefootmark{d}  & $-70$ & $-455$ & 184 & 58 & 170$\pm$9 & $\sim290$\tablefootmark{g} \\
IRAS~F18216+6419 & 0.2974\tablefootmark{a} & Sy/QSO & 6.9\tablefootmark{d} & $-$ & $-$ & $-$ & $-$ & $<18$\tablefootmark{f} & $-$\\
\hline
J131911+095051 & 6.1330\tablefootmark{c} & QSO & 25.1\tablefootmark{e} & $-154$ & $-314$ & 38.4 & 14.4 & 0.14$\pm0.048$ & $\sim200$\tablefootmark{g} \\
\hline
\end{tabular}
\tablefoot{
\tablefoottext{a}{Redshift from \cite{rhc_wang09}.}
\tablefoottext{b}{Redshift from \cite{rhc_rupke05}.}
\tablefoottext{c}{Redshift from \cite{rhc_wang13}.}
\tablefoottext{d}{$L_{\rm FIR}$ from \cite{rhc_wang09}.}
\tablefoottext{e}{Calculated from the modeling of the spectral energy distribution (see Appendix~\ref{QSO_SED})}
\tablefoottext{f}{3$\sigma$ upper limit assuming a linewidth of 500~km~s$^{-1}$}
\tablefoottext{g}{Calculated using Equation~\ref{eq:EW}.}
\tablefoottext{h}{This value corresponds to an inflow rate (see Section~\ref{sec:inflow}).}
}
\end{center}
\end{table*}

In this Letter we extend the search for molecular inflows and outflows based on the \OH\ doublet into two relatively unexplored and interesting regimes: (1) ultraluminous infrared galaxies (ULIRGs; $L_{\rm IR}>10^{12}~L_{\odot}$) at $z\sim0.2-0.3$, which are more luminous and at higher redshift than the rest of the {\it Herschel}/PACS OH database with the exception of the sample in \cite{rhc_calderon16}. At this redshift we expect these systems to have higher molecular gas fractions than nearby ULIRGs \citep[e.g.,][]{rhc_tacconi18}. (2) QSOs at $z\sim6$, where we would like to test if large scale molecular outflows play a role in shutting  down star formation and black hole growth in the galaxy. Also, OH can prove to be an effective tool to discover molecular outflows in high redshift QSOs. To date, and with the exception of QSO~J1148+5251 \citep{rhc_maiolino12,rhc_cicone15}, there are no molecular outflows detected in individual CO and/or \cii\ spectra of $z\sim6$ QSOs \citep[e.g.,][]{rhc_wang13,rhc_wang16,rhc_willott15,rhc_shao17,rhc_decarli17,rhc_decarli18,rhc_decarli19}.  

\section{Sample and observations}\label{sec:sample}

Our study focusses on five systems. Four of them are ULIRGs at intermediate redshift ($0.19<z<0.29$) selected from The Imperial IRAS-FSC Redshift Catalogue \citep{rhc_wang09}, and the fifth one is \qsoname, a QSO at $z=6.13$ \citep{rhc_wang13,rhc_shao17}. The redshifts, infrared luminosities, and classifications of these galaxies are listed in Table~\ref{tab:sample}. As shown in Figure~\ref{fig:sample}, these galaxies contribute to double the number of galaxies with \OH\ observations in the $\sim$0.2-0.3 redshift range, and push the search for outflows based on \OH\ to $z\sim6$. Based on a WISE color-color diagram (Figure~\ref{fig:sample}), three of the ULIRGs in our sample have color consistent with Seyfert/QSO systems following the criteria by \cite{rhc_wright10} \cite[see also][]{rhc_mateos12,rhc_assef13}. IRAS~F10091+4704, on the other hand, appears to be a starburst-dominated system as suggested by its WISE and mid-infrared colors \citep{rhc_veilleux09}.

Observations of the \OH\ doublet for the four ULIRGs were carried out using the PACS instrument  \citep{rhc_poglitsch10} on-board the {\it Herschel} Space Observatory \citep{rhc_pilbratt10} as part of a Herschel Open Time program OT2\_jgracia\_1 (PI, J. Graci\'a-Carpio). The observation IDs following the list order in Table~\ref{tab:sample} are 1342270673, 1342270662, 1342268181, and 1342270020. The data were reduced using the PACS data reduction and calibration pipeline included in HIPE \citep[Herschel Interactive Processing Environment;][]{rhc_ott10}. The spectra used in the analysis were extracted from the central spaxel of $9.4\arcsec\times9.4\arcsec$ in size, and the systemic velocity was set based on the known spectroscopic redshifts listed in Table~\ref{tab:sample}.

The $z=6.13$ QSO \qsoname\ was observed with ALMA Band~7  as part of Project 2012.1.00391.S (PI J. Graci\'a-Carpio). In order to cover a continuous velocity range of approximately $\pm2500$~km~s$^{-1}$ around the systemic velocity where we expected the \OH\ transition \citep[based on $z_{\rm [CII]}=6.1330\pm0.0007$;][]{rhc_wang13}, we observed the source using two independent observing blocks and then stitched together pairs of consecutive spectral windows from each block to continuously cover the [348.18,355.60]~GHz frequency range. The first and second halves of the frequency range were observed for 41 and 80~min with 32 antennas, reaching sensitivities of 0.19 and 0.14~mJy~beam$^{-1}$
 in 150~km~s$^{-1}$ velocity bins, respectively. The half-power beam width (HPBW) in each observing block is approximately the same, ${\rm HPBW}=0.63\arcsec \times 0.44\arcsec$. 

\begin{figure*}
\centering
\includegraphics[scale=0.15]{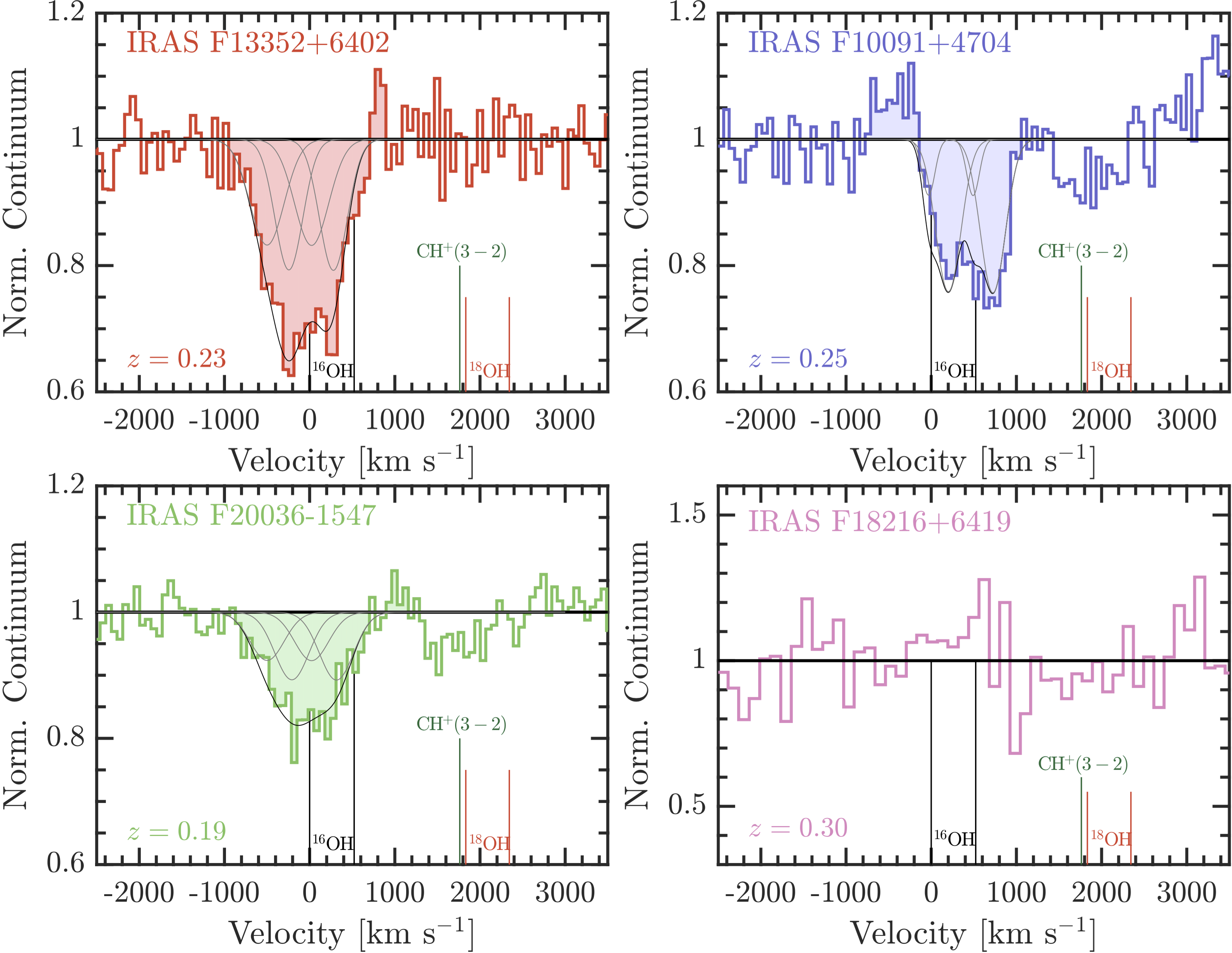}
\caption{Observed PACS spectra (continuum-normalized) of the OH transition at 119~$\mu$m of IRAS F13352+6402 (top-left), IRAS F10091+4704 (top right), IRAS F20036-1547 (bottom-left), and IRAS F18216+6419 (bottom-right). The origin of the velocity scale corresponds to the OH 119.23~$\mu$m transition at the systemic velocity. The vertical lines mark the positions of the $^{16}$OH and $^{18}$OH doublets and the CH$^{+}(3-2)$ transition. Two of the systems, IRAS F20036-1547 and IRAS F13352+6402, show evidence of fast (maximum velocity $\sim1,000$~km~s$^{-1}$) molecular outflowing gas, while IRAS F10091+4704 show evidence for a molecular gas inflow. In the case of QSO IRAS F18216+6419 we do not significantly detect any of the OH features neither in absorption nor in emission.
 }\label{fig:int_spec}
\end{figure*}

\section{Results}

\subsection{ULIRGs at $z\sim0.2-0.3$}

The spectra of the ULIRGs extracted from the {\it Herschel}/PACS data are shown in Figure~\ref{fig:int_spec}. The vertical lines mark the positions of the $^{16}$OH and $^{18}$OH doublets and the CH$^{+}(3-2)$ line. The second independent absorption feature detected in IRAS~F10091+4704~appears to be associated to the CH$^{+}(3-2)$ transition rather than the $^{18}$OH doublet, similar to the case of NGC~4418 \citep{rhc_gonzalez-alfonso12}. Following a procedure similar to that used by \cite{rhc_veilleux13} and \cite{rhc_stone16}, we use the code {\it PySpecKit} \citep{rhc_ginsburg11} to fit the $^{16}$OH~($\lambda$119.233,119.441~$\mu$m) doublet using four Gaussian components (two components for each line of the doublet), each characterized by their amplitude, peak position, and standard deviation. From the Gaussian fits we measured $v_{50}$ and $v_{84}$, i.e., the velocities above 50\%, and 84\% of the  absorption takes place, respectively. The results are listed in Table~\ref{tab:sample}, including also the equivalent width (EW), the EW of gas with $v<-200~{\rm km~s}^{-1}$ (i.e., excluding OH gas at rest), and the flux measured in the \OH\ doublet. We estimate an overall error of $\pm50$~km~s$^{-1}$ for the median velocities. We then follow the conservative criteria by \cite{rhc_rupke05b} to classify the observed gas flows as inflows if $v_{50}\gtrsim+50$~km~s$^{-1}$ or outflows if $v_{50}\lesssim-50$~km~s$^{-1}$. According to this convention, IRAS F13352+6402 ($v_{50}=-115$~km~s$^{-1}$) and IRAS~F20036-1547 ($v_{50}=-70$~km~s$^{-1}$) are considered outflows, while IRAS~F10091+4704 ($v_{50}=+430$~km~s$^{-1}$) is classified as an inflow. IRAS~F18216+6419, on the other hand, has no detection of the OH doublets neither in emission nor in absorption. In this case, however, the $1\sigma$ uncertainty in the normalized continuum is of the order of $7\%$ in 60~km~s$^{-1}$ channels, about a factor of $\sim2$ worse than in the other ULIRGs.

\subsection{QSO \qsoname at $z=6.13$ }

\begin{figure}[]
\begin{center}
\includegraphics[scale=0.08]{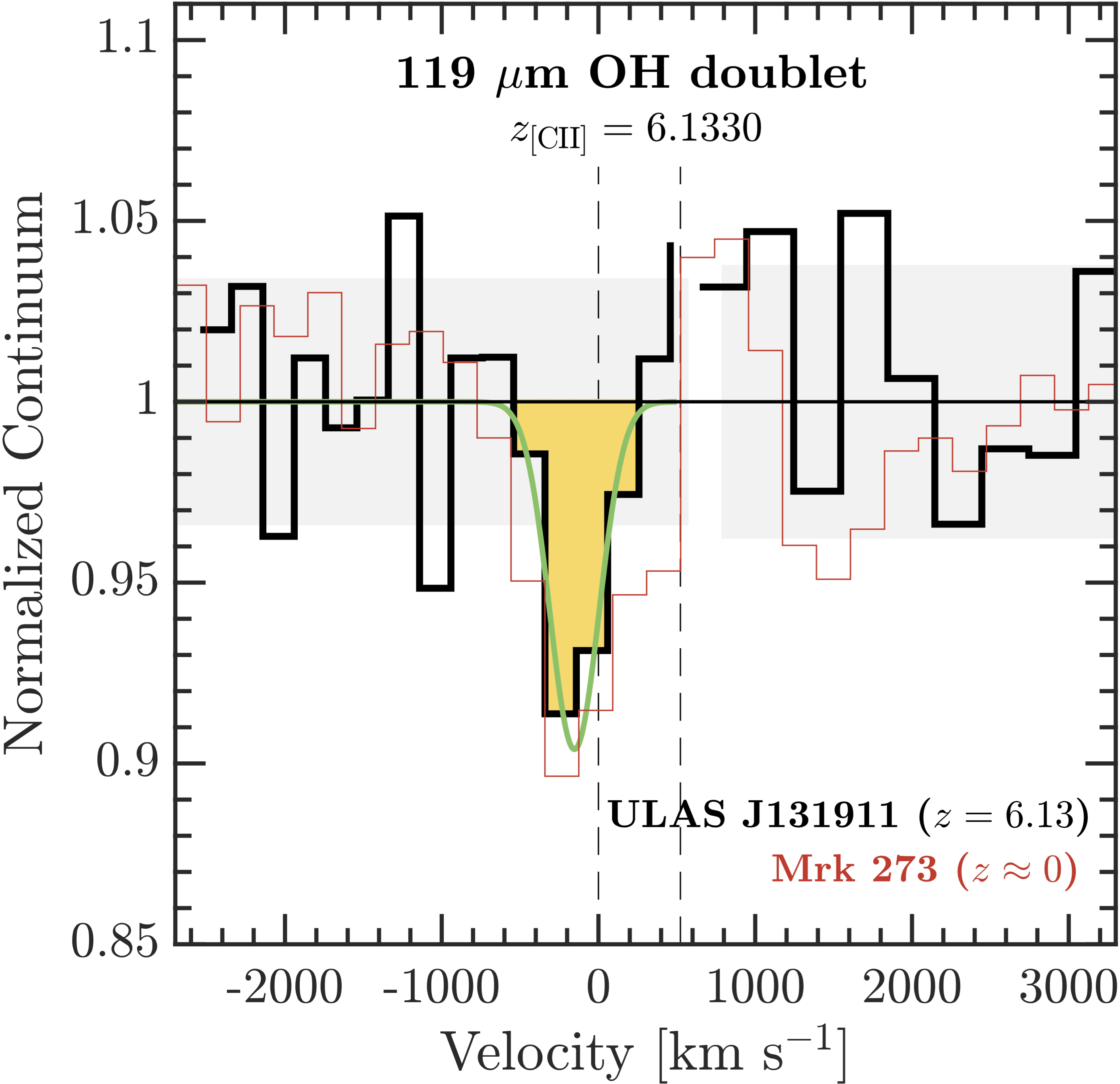}
\caption{OH 119~$\mu$m spectrum of \qsoname\ (black and yellow). The shaded grey rectangles show the  $\pm1\sigma$ uncertainty on the continuum. The systemic velocity is set by the observed redshift of the source based on a secure detection of the \cii\ line \citep{rhc_wang13}. The vertical dashed lines show the expected position for the OH~119~$\mu$m doublet. The green line shows the best single-component Gaussian fit to the data. For comparison, we include the {\it Herschel}/PACS OH 119~$\mu$m spectrum of local ULIRG Mrk~273 (red) smoothed in velocity to match the resolution of the ALMA OH spectrum.}\label{fig:qso_spec}
\end{center}
\end{figure}

Figure~\ref{fig:qso_spec} shows the ALMA OH spectrum extracted from the inner $\sim3$~kpc region of \qsoname. The rest-frame velocity is set based on a precise redshift measurement derived from the detection of the \cii\ line by \cite{rhc_wang13}. We observe a tentative absorption feature blueshifted from the systemic velocity by $\sim150$~km~s$^{-1}$ that could indicate the presence of a molecular outflow. The significance of the detection ---based on the integrated flux inside a single Gaussian fit--- is $\approx3\sigma$, and the median ($v_{50}$) and $v_{84}$ velocities are $-154$~km~s$^{-1}$ and $-314$~km~s$^{-1}$, respectively. As Figure~\ref{fig:qso_spec} shows, the potential outflow signature in \qsoname\ resembles in intensity and velocity structure (smoothed to match the ALMA spectrum velocity resolution) the \OH\ absorption profile detected in the nearby ULIRG Mrk~273 \citep{rhc_veilleux13}. \footnote{We use Mrk 273 for the comparison as this system has an \OH\ absorption feature median velocity  ($v_{50}=-201$~km~s$^{-1}$) and a continuum-normalized peak absorption intensity ($I_{\rm peak}/I_{\rm cont}\approx0.9$) similar to those of the outflow in \qsoname.}

In Appendix~\ref{QSO_SED} we discuss the ALMA Band~7 dust continuum properties of the QSO host, including a refined measurement of the far-infrared luminosity and the dust mass.

\section{Discussion}

\subsection{Molecular Outflows}

\begin{figure*}[ht!]
\begin{center}
\includegraphics[scale=0.15]{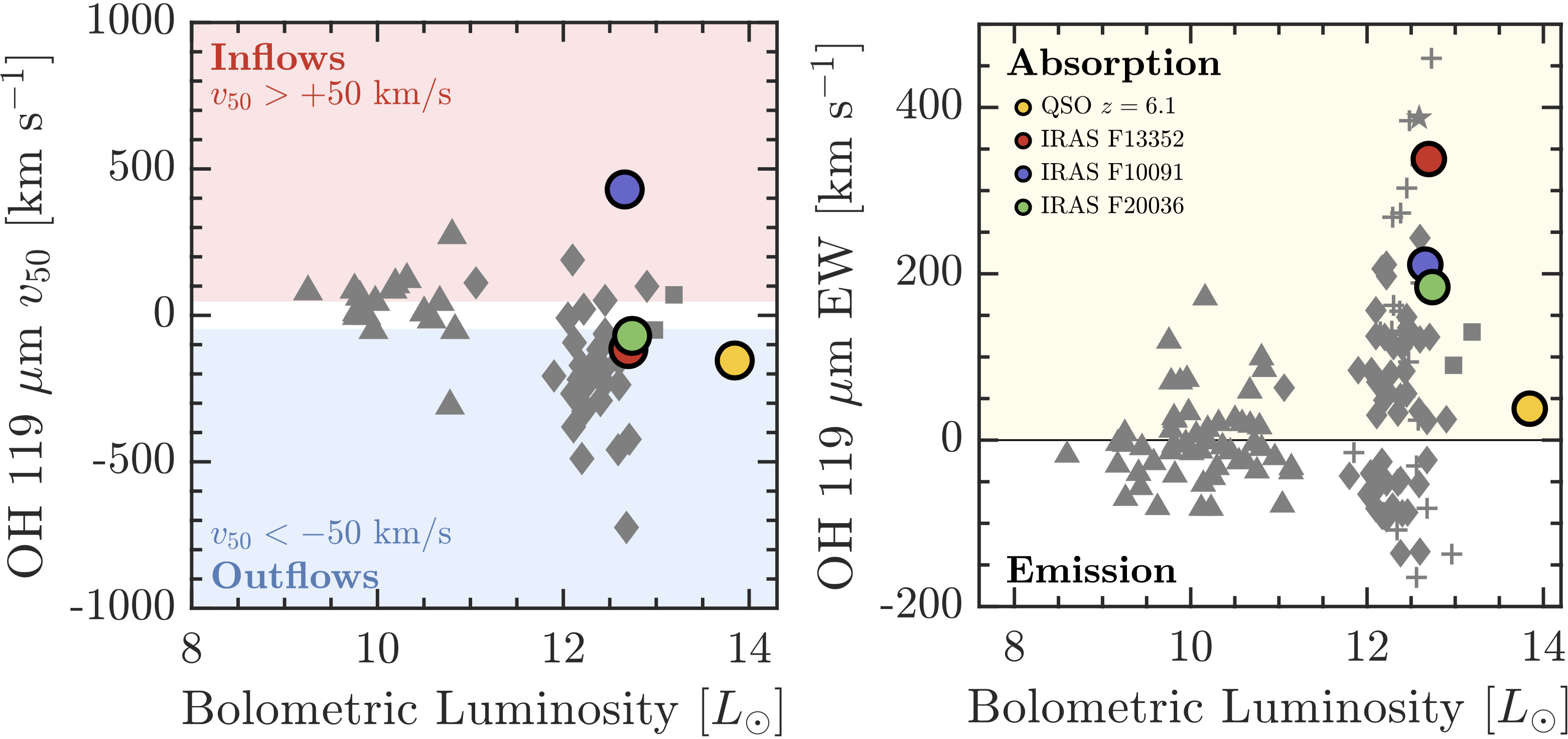}
\caption{{(\it Left)} \OH\ line median velocities ($v_{50}$) as a function of bolometric luminosity. Following the criteria by \cite{rhc_rupke05b} we classify gas with median velocity $v_{50}>+50$~km~s$^{-1}$ as inflows (top, red part of the panel) and gas with $v_{50}<-50$~km~s$^{-1}$ as outflows (bottom, blue part of the panel). Each galaxy is represented by a symbol: those in our sample are shown with color circles, and additional samples of nearby AGNs, (U)LIRGs and QSOs, and $z\sim0.3$ HyLIRGs are shown as triangles, diamonds and squares, respectively. The \OH\ line median velocities and bolometric luminosities of these galaxies are drawn from \cite{rhc_stone16}, \cite{rhc_veilleux13}, and \cite{rhc_calderon16}. {(\it Right)} \OH\ equivalent width as a function of bolometric luminosity. The samples of galaxies and the symbols are the same as the left panel with the addition of the sample of ULIRGs in \cite{rhc_spoon13} shown with crosses.}\label{fig:comp}
\end{center}
\end{figure*}

Figure~\ref{fig:comp} shows the median velocities and equivalent widths of the \OH\ profiles measured in IRAS~F13352+6402, IRAS~F20036-1547, IRAS~F10091+4704, and QSO~J131911+095051. We also include measurements of nearby AGNs \citep{rhc_stone16}, ULIRGs and QSOs \citep{rhc_veilleux13,rhc_spoon13}, and $z\sim0.3$ hyperluminous infrared
galaxies \citep[$L_{\rm IR}>10^{13}$~L$_{\odot}$, HyLIRGs; ][]{rhc_calderon16}. The wide spread in \OH\ EWs and median velocities at high bolometric luminosities could be partly explained by evolutionary effects, as the covering fraction of the 119~$\mu$m continuum by the outflow is expected to vary as systems transition from extremely buried phases with no wide-angle outflows to phases where the outflow activity has subsided \cite[e.g.,][]{rhc_veilleux13,rhc_gonzalez-alfonso17,rhc_falstad19}. These samples combined show that above a certain luminosity threshold (${\rm log}_{10}(L_{\rm bol}/L_{\odot})\gtrsim11.5$), galaxies tend to have higher outflow  (blueshifted) OH velocities as a function of increasing bolometric luminosity \citep[for a detailed discussion, see][]{rhc_stone16}. 
The three ULIRGs in our sample stand out as having large \OH\ equivalent widths relative to other systems of similar luminosity. In particular, IRAS~F13352+6402 has one of the highest equivalent widths measured to date, only comparable to that measured in Arp~220 \citep{rhc_gonzalez-alfonso12} or IRAS~20087-0308 \citep{rhc_spoon13}.

Calculating the mass outflow rate solely based on the often optically thick \OH\ doublet profile is extremely challenging. Models that constrain the structure and energetics of the outflow, like the one by \cite{rhc_gonzalez-alfonso14,rhc_gonzalez-alfonso17}, require the simultaneous detection of multiple OH transitions (e.g., OH~65~$\mu$m, 79~$\mu$m, 84~$\mu$m, and 119~$\mu$m), as these are formed at different locations in the wind depending on the excitation conditions of the gas that is radiatively pumped in the outflow. Given that we only have a single OH transition available, we need an alternative method to  estimate the mass outflow rate. One possibility is to use an empirical relation between the observed  EW of the \OH\ absorption component of the profile, the far-infrared luminosity and the mass outflow rate derived from the model by \cite{rhc_gonzalez-alfonso17} (see Appendix~\ref{app:ew} for details).\footnote{A similar method is used by \cite{rhc_spilker18} to estimate the mass outflow rate solely based on the \OH\ doublet in a lensed, $z=5.3$ submillimiter galaxy. However,  \cite{rhc_spilker18} do not scale the \OH\ equivalent width by $L_{\rm FIR}^{1/2}$.} If IRAS~F20036-1547 and IRAS~F13352+6402 follow the trend observed in nearby ULIRGs shown in Figure~\ref{fig:EW_dMdt}, then following Equation~\ref{eq:EW} we expect the mass outflow rates to be $\dot{M}_{\rm out,mol}\sim290~M_{\odot}~{\rm yr}^{-1}$ and $\sim570~M_{\odot}~{\rm yr}^{-1}$, respectively.

Both IRAS~F20036-1547 and IRAS~F13352+6402 are classified as AGN according to their WISE colors, which implies that the outflows could be driven by a combination of AGN and starburst activity. For IRAS~F13352+6402 the AGN contribution to the bolometric luminosity may be as low as $\sim10\%$ \citep{rhc_nardini10}, so if we convert the far-infrared luminosity associated to the starburst into a SFR we find a mass loading factor $\eta\gtrsim1.5$. For IRAS~F20036-1547 there are no constraints on the AGN contribution to the luminosity, so by assuming that all the far-infrared emission is associated to the starburst we estimate a mass loading factor lower limit of  $\eta\gtrsim0.7$. These mass loading factors are consistent with outflows driven purely by star formation \citep[e.g., $\eta\sim1-3$ in NGC~253 or M82;][]{rhc_bolatto13,rhc_leroy15} or mixed AGN-starburst activity \citep[e.g., $\eta\sim1-3$ in IRASF~10565+2448 or Mrk~273;][]{rhc_cicone14}, so additional information is needed to identify the dominant source behind the wind.

In terms of outflow incidence, if we combine our sample of 4 ULIRGs at $z\sim0.2-0.3$ with the sample of 5 HyLIRGs at $z\sim0.3$ from \cite{rhc_calderon16}, we estimate an OH-based outflow detection rate in intermediate redshift ULIRGs of $\sim45\%$ (4/9). This fraction is higher than the modest detection rate found among Local Volume Seyferts \citep[$\sim25\%$;][]{rhc_stone16}, but lower than the  detection fraction measured in {$z<0.2$} (U)LIRGs \citep[$\sim70\%$;][]{rhc_veilleux13}.

\subsection{Molecular inflows}\label{sec:inflow}

Molecular gas inflows towards the central region in interacting/merging galaxies are expected to periodically replenish the molecular gas that is consumed by star formation, the super massive black hole and/or ejected by outflows. In the case of IRAS~F10091+4704 --a merger system where the nuclei have
apparently coalesced \citep{rhc_veilleux02}--, we detect a clear inverted P-Cygni profile which provides strong evidence for the presence of a molecular gas inflow. This profile is unique among ULIRGs, and only comparable in significance to the inverted \OH\ P-Cygni profile observed in Circinus \citep{rhc_stone16}. As Figure~\ref{fig:comp} shows, the inflowing molecular gas in this galaxy has the highest OH median velocity ($v_{50}=+430$~km~s$^{-1}$) measured to date. If we use a scaled version the OH-based mass inflow rate measured in NGC~4418 of $12~M_{\odot}~{\rm yr}^{-1}$ \citep{rhc_gonzalez-alfonso12}, we estimate a molecular inflow rate for IRAS~F10091+4704 of $\dot{M}_{\rm mol,in}\sim12 \times(211/113)\times(3.7\times10^{12}/2\times10^{11})^{1/2}\sim100~M_{\odot}~{\rm yr}^{-1}$. This inflow rate is about 30\% of the total SFR of the system.

In terms of demographics, the inflow in IRAS~F10091+4704 is the only one detected in the combined sample of 9 intermediate redshift ULIRGs from this work and \cite{rhc_calderon16}. This $\sim10\%$ inflow detection rate is similar to that measured in (U)LIRGs by \cite{rhc_veilleux13} using the \OH\ doublet and \cite{rhc_rupke05} using the \nad\ line in absorption. The observed low inflow detection rate has been interpreted as a consequence of the planar or filamentary geometry of inflows that subtend a much smaller covering angle than outflows. Another explanation could be that fast winds may disturb the molecular gas, preventing it from inflowing to the center. Perhaps the lack of strong AGN activity in IRAS~F10091+4704 means that the system is going through a phase of efficient gas replenishment of its nuclear region due to the absence of powerful AGN-driven winds.

\subsection{Evidence for a molecular outflow in a $z\sim6$ QSO}

With the exception of QSO J1148+5251 \citep{rhc_maiolino12,rhc_cicone15}, there are no other individual QSOs at $z\sim6$ with direct detections of molecular outflows. Part of the problem is that observations of most of these high-$z$ QSOs are not deep enough to detect broad wings of emission in their \cii\ or CO spectra \citep[for a discussion see][]{rhc_bischetti18,rhc_stanley19}. In the case of \qsoname, \cite{rhc_wang13} and \cite{rhc_shao17} failed to detect any outflow signature in the \cii\ spectra. The fact that we tentatively detect the molecular outflow in the \OH\ spectra with comparable sensitivity than the \cii\ observations makes the OH transitions an interesting alternative to search for outflows in high-$z$ QSOs.

Based on the empirical relation between $\dot{M}_{\rm out,mol}$, EW$^{v<-200}_{\rm OH}$, and $L_{\rm FIR}$ (Appendix~\ref{app:ew}) we determine a molecular mass outflow rate for \qsoname\ of $\dot{M}_{\rm out,mol}\sim200~M_{\odot}~{\rm yr}^{-1}$. 
This value is only half the star formation rate of the QSO (${\rm SFR}=375~M_{\odot}~{\rm yr}^{-1}$ if we assume that only 10\% of $L_{\rm FIR}$ is associated with the starburst), which is in stark contrast with the typical molecular mass loading factor of $\eta\sim1-10$ measured in local (U)LIRGs \citep[e.g.,][]{rhc_veilleux13,rhc_gonzalez-alfonso17,rhc_rhc19b}. The modest mass outflow rate in \qsoname\ is, however, comparable to the low values found in the stacking analysis of the \cii\ spectra of $z\sim5-7$ QSOs by \cite{rhc_bischetti18} ($\dot{M}_{\rm out,mol}\sim100-200~M_{\odot}~{\rm yr}^{-1}$) and \cite{rhc_stanley19}.

The kinetic power in the molecular outflow of \qsoname\ ---calculated as $P_{\rm K,mol}=1/2\times\dot{M}_{\rm out,mol}\times v_{84}^2$--- is less than 0.1\% of the bolometric luminosity. For an energy-driven wind the expectation is $\sim5\%$ if the coupling between the outflow and the ISM is 100\% \citep{rhc_f-g12,rhc_costa14,rhc_zubovas14b}. This result, combined with the modest molecular mass outflow rate, suggests that AGN-driven winds already cleared multiple paths through the nuclear ambient medium of \qsoname, reducing the efficiency of the ejective or mechanical mode of negative feedback. 

We note, however, that there are a couple of alternatives that could increase the total mass outflow rate in \qsoname\ and modify this interpretation. For example, it could be that the molecular gas in the outflow transitions to the atomic phase as it moves further away from the host \citep[e.g., the case of M~82;][]{rhc_leroy15}, or that the outflow has a diffuse molecular component that extends beyond the nuclear region and thus can only be detected by other tracers such as CO, \ci, or \cii\ line emission. Certainly, deeper and higher angular resolution ALMA observations are required to confirm the existence of the molecular outflow in \qsoname\ and better characterize it properties. 

\section{Conclusions}

Based on ALMA and {\it Herschel}/PACS observations of the \OH\ doublet we searched for inflows and outflows of molecular gas in four ULIRGs at $z\sim0.2-0.3$ and one QSO at $z\sim6$. We detect powerful molecular outflows in two ULIRGs --IRAS~F20036-1547 and IRAS~F13352+6402--, with molecular mass outflow rates as high as their star formation rates, consistent with expectations from quenching models by way of ejective feedback. We also detect an inverted P-Cygni profile in the starburst-dominated system IRAS~F10091+4704 which is unique among ULIRGs and suggests a strong inflow of molecular gas ($\dot{M}_{\rm mol,in}\sim100~M_{\odot}~{\rm yr}^{-1}$). Finally, we tentatively detect ($\approx3\sigma$) a molecular outflow in the QSO \qsoname\ at $z=6.13$. The molecular mass outflow rate is modest ($\dot{M}_{\rm mol,out}\sim200~M_{\odot}~{\rm yr}^{-1}$), consistent with constraints obtained from stacking of the \cii\ line in QSOs at $z\sim6$ \citep{rhc_bischetti18,rhc_stanley19}, and suggest that perhaps we are witnessing a phase in the QSO evolution where mechanical feedback have subsided.

\begin{acknowledgements}
     We thank the referee for useful comments and suggestions that improved this Letter. R. H-C. would like to dedicate this letter to Olivia on her third birthday. 
     E.GA is a Research Associate at the Harvard-Smithsonian Center for Astrophysics, and thanks the Spanish Ministerio de Econom\'{\i}a y Competitividad for support under project ESP2017-86582-C4-1-R. R.M. acknowledges ERC Advanced Grant 695671 "QUENCH?. This paper makes use of the following ALMA data: ADS/JAO.ALMA\#2012.1.00391.S. ALMA is a partnership of ESO (representing its member states), NSF (USA) and NINS (Japan), together with NRC (Canada), MOST and ASIAA (Taiwan), and KASI (Republic of Korea), in cooperation with the Republic of Chile. The Joint ALMA Observatory is operated by ESO, AUI/NRAO and NAOJ. PACS has been developed by a consortium of institutes led by MPE (Germany) and including UVIE (Austria); KU Leuven, CSL, IMEC (Belgium); CEA, LAM (France); MPIA (Germany); INAF-IFSI/OAA/OAP/OAT, LENS, SISSA (Italy); IAC (Spain). This development has been supported by the funding agencies BMVIT (Austria), ESA-PRODEX (Belgium), CEA/CNES (France), DLR (Germany), ASI/INAF (Italy), and CICYT/MCYT (Spain). PACS has been developed by a consortium of institutes led by MPE (Germany) and including UVIE (Austria); KU Leuven, CSL, IMEC (Belgium); CEA, LAM (France); MPIA (Germany); INAF-IFSI/OAA/OAP/OAT, LENS, SISSA (Italy); IAC (Spain). This development has been supported by the funding agencies BMVIT (Austria), ESA-PRODEX (Belgium), CEA/CNES (France), DLR (Germany), ASI/INAF (Italy), and CICYT/MCYT (Spain).This research made use of {\it PySpecKit}, an open-source spectroscopic toolkit hosted at http://pyspeckit.bitbucket.org.
\end{acknowledgements}

\begin{appendix}

\section{The sample and WISE color-color classification}

\begin{figure*}[b!]
\begin{center}
\includegraphics[scale=0.15]{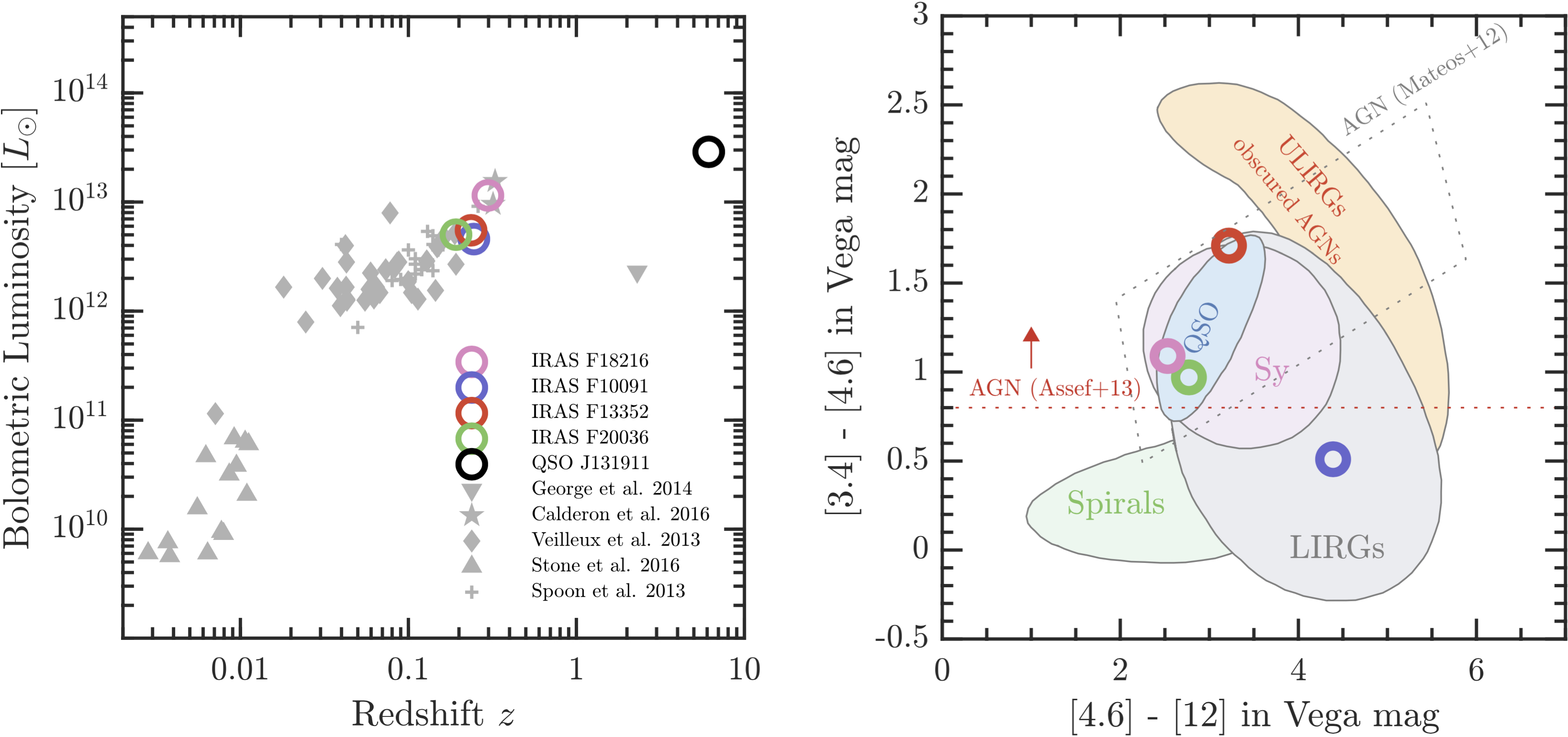}
\caption{({\it Left}) Bolometric luminosity of galaxies with molecular outflows or inflows observed in the OH119 doublet as a function of redshift. The ultraluminous infrared galaxies (ULIRGs) and high-$z$ QSO in our sample are shown as color circles: IRAS F18216+6419, IRAS F10091+4704, IRAS F13352+6402, IRAS F20036+1547, and ULASJ131911. For comparison of the parameter space covered by our sample we include HyLIRGs from \cite{rhc_calderon16}, ULIRGs and QSO from \cite{rhc_veilleux13} and \cite{rhc_spoon13}, Seyferts from \cite{rhc_stone16}, and one SMG (The Eyelash) at $z\sim2$ from \cite{rhc_george14}. ({\it Right}) WISE color-color diagram of the ultraluminous infrared galaxies in our sample. The background illustrates the typical locations of different classes of objects \citep{rhc_wright10}. The dashed grey box \citep{rhc_mateos12} and the red horizontal line \citep{rhc_assef13} shows different criteria for AGN classification. According to this, three of our systems are classified as AGN, while IRAS F10091+4704 can be considered a starburst-dominated system.}\label{fig:sample}
\end{center}
\end{figure*}

Figure~\ref{fig:sample} shows the redshift and bolometric luminosities of the sample and a WISE color-color diagram used to discriminate between AGN or starburst dominated systems following the criteria by \cite{rhc_wright10}, \cite{rhc_mateos12}, and \cite{rhc_assef13}.

\section{ALMA dust continuum map and updated SED}\label{QSO_SED}

\begin{figure*}[ht!]
\begin{center}
\includegraphics[scale=0.052]{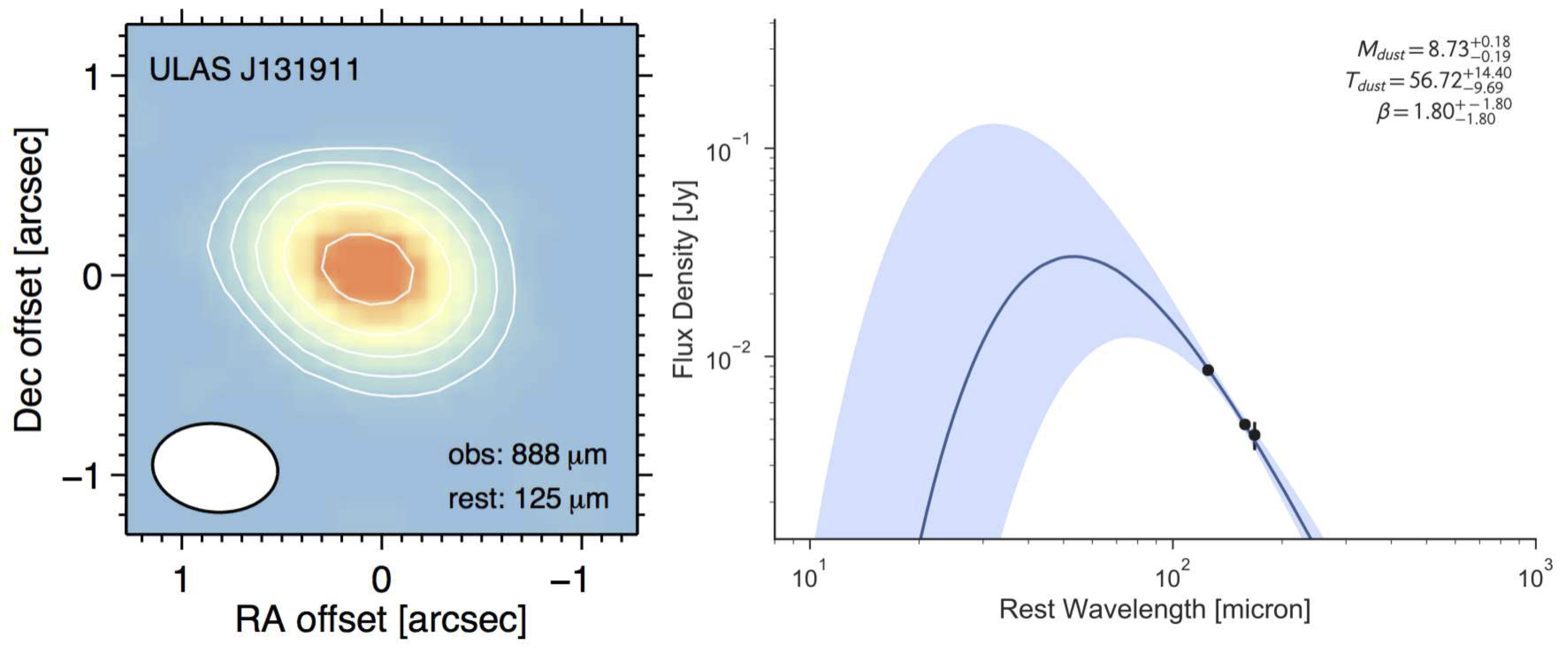}
\caption{{\it (Left)} Dust continuum map at observed 888~$\mu$m (rest-frame 125~$\mu$m) of the QSO ULAS J131911. For a redshift of $z=6.1330$ \citep[based on a ${\rm [CII]}~158~\mu$m detection by][]{rhc_wang13} this corresponds to a rest-frame wavelength of 124.5~$\mu$m. The white contours corresponds to $S/N = 5$, 10, 20, 40, and 80~$\sigma$. The synthesized beam ($\theta=0.63\arcsec\times0.44\arcsec$) is shown in the lower left corner. {\it (Right)} Dust continuum SED combining photometric measurements at rest-frame wavelengths 125~$\mu$m (our work), 158~$\mu$m \citep{rhc_shao17}, and 168~$\mu$m \citep{rhc_wang13}. The solid blue line corresponds to the best modified blackbody fit to the data. This yield a dust temperature of $T_{\rm dust}=56.7$~K and dust mass of $M_{\rm dust}=10^{8.73}$~$M_{\odot}$. The shaded region indicates the 2.5 and 97.5 percentile for the spread of model SEDs based on the posterior distributions for the parameters. }\label{fig:qso_cont}
\end {center}
\end{figure*}

Based on the ALMA Band~7 part of the spectrum that is free from OH features we construct a dust continuum map of the QSO at a rest-frame wavelength of 125~$\mu$m (observed wavelength of 888~$\mu$m). As Figure~\ref{fig:qso_cont} shows, dust emission in the QSO is detected with high S/N ($\gtrsim70$) and has a size (FWHM deconvolved from the beam) of $2.5\times1.5$~kpc ($0.44\arcsec \times 0.25\arcsec$). We measure a flux density of $S_{\rm 125~\mu m, rest}=8.56\pm0.12$~mJy, which combined with previous dust continuum measurements at rest-frame wavelengths 158~$\mu$m \citep{rhc_shao17} and 168~$\mu$m \citep{rhc_wang13} helped us to improve the constraints on the dust mass, dust temperature, and the total infrared luminosity. Figure~\ref{fig:qso_cont} (right) shows the observed spectral energy distribution and the best modified blackbody fit. The latter yields a FIR luminosity of  log$_{10}(L_{\rm FIR}/L_{\odot})=13.4\pm0.16$, dust temperature of $T_{\rm dust}=56.7^{+14.40}_{-9.69}$~K and a dust mass of log$_{10}(M_{\rm dust}/M_{\odot})=8.73^{+0.18}_{-0.19}$. The molecular gas mass based on a CO(6-5) line detection, assuming a CO excitation ladder similar to SDSS J114816.64+525150.3 \citep{rhc_riechers09}, is $M_{\rm mol}=1.5\times10^{10}~M_{\odot}$ \citep{rhc_wang13}. This results in a dust-to-gas mass ratio of $M_{\rm dust}/M_{\rm mol}=0.036$. 

The dust-to-gas mass ratio in \qsoname\ is comparable to those values measured in nearby star-forming galaxies \citep[e.g.,][]{rhc_draine07b}, other $z\sim6$ QSOs \citep[e.g.,][]{rhc_calura14}, and about two orders of magnitude higher than the dust-to-gas mass ratio measured in local low-metallicity galaxies ---often considered as analogs of the dominant population of primeval systems in the early Universe \citep[e.g.,][]{rhc_rhc12,rhc_fisher14,rhc_remy14}.

\section{\OH\ equivalent width -- mass outflow rate empirical relation}\label{app:ew}

\begin{figure*}[ht!]
\begin{center}
\includegraphics[scale=0.1]{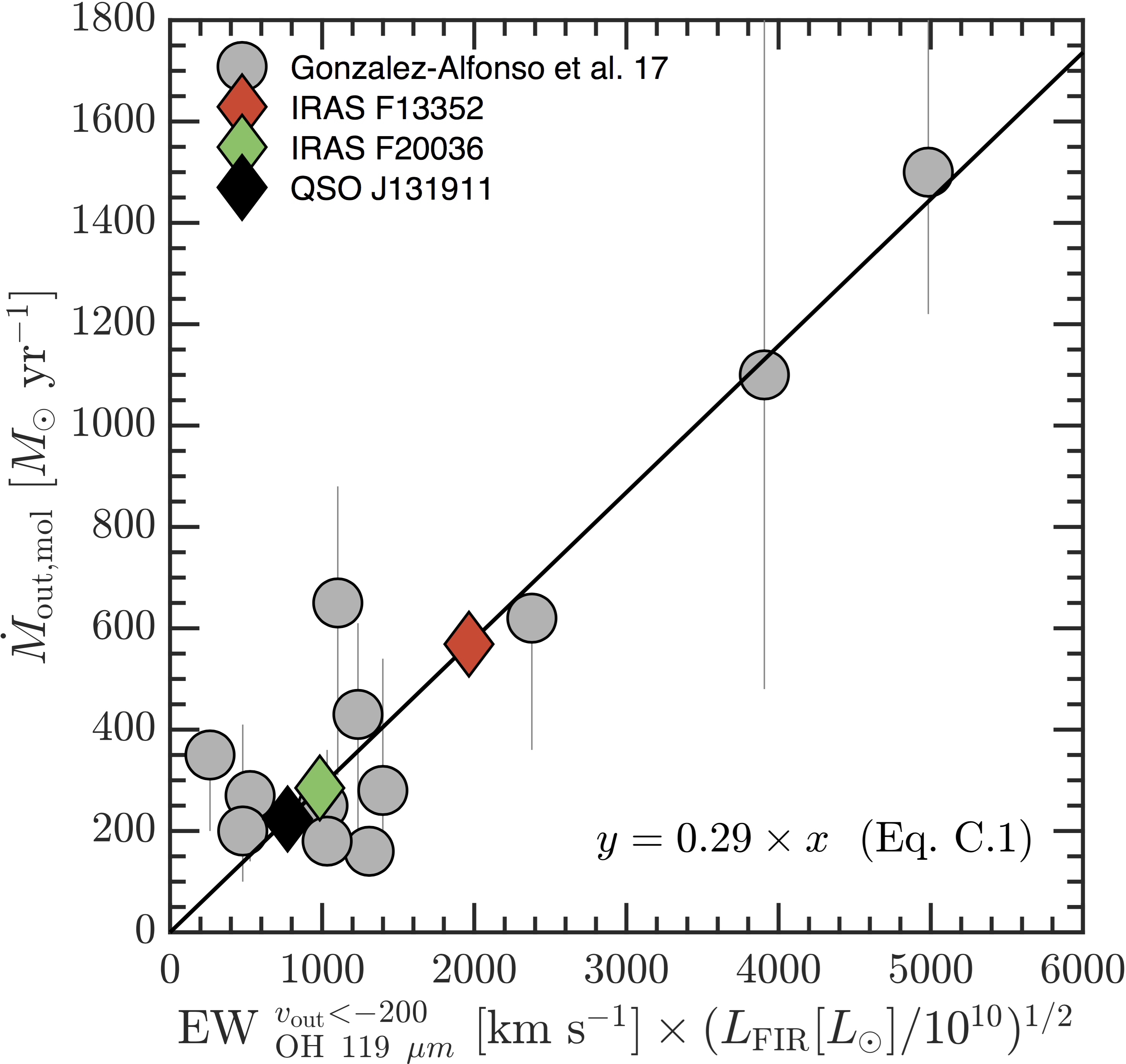}
\caption{Total molecular mass loss rates inferred from models of multiple OH transitions in 12 local ULIRGs by \citet[][; gray circles]{rhc_gonzalez-alfonso17} as a function of the observed \OH\ equivalent width for gas with velocities $v<-200$~km~s$^{-1}$ (${\rm EW}^{~v<-200~{\rm km~s^{-1}}}_{\rm OH~119~\mu m}$) and the square root of the FIR luminosity. The implied mass outflow rates for IRAS~F13352+6402, IRAS~F20036-1547, and QSO~J131911+095051 are shown as red, green, and black diamonds, respectively.}
\label{fig:EW_dMdt}
\end {center}
\end{figure*}

The models of \cite{rhc_gonzalez-alfonso17} are the result of a combination of a core component, which describes the continuum and line absorption produced in the nuclear region of galaxies, and an envelope, which consist of an expanding shell of gas and dust. The molecular mass outflow rate ($\dot{M}_{\rm mol,out}$) is proportional to the OH column density and the outflow size, which in turn are proportional to the OH equivalent width (${\rm EW}_{119~{\rm \mu m}}$) and the square root of the infrared luminosity ($L_{\rm FIR}^{1/2}$), respectively, provided that the outflows remain self-similar with increasing $L_{\rm FIR}$. To derive an empirical relation that connects $\dot{M}_{\rm mol,out}$ to the observed ${\rm EW}_{119~{\rm \mu m}}$ and $L_{\rm FIR}$, we first recalculate the equivalent widths of the galaxies in the study of \cite{rhc_gonzalez-alfonso17} to only consider gas with velocities more blueshifted than $-200$~km~s$^{-1}$, i.e., to exclude OH gas at rest \cite[see also, e.g.,][]{rhc_spoon13}. We refer to this quantity as ${\rm EW}^{~v<-200~{\rm km~s^{-1}}}_{\rm OH~119~\mu m}$. Figure~\ref{fig:EW_dMdt} shows the relation between $\dot{M}_{\rm mol,out}$ --inferred from the models of \cite{rhc_gonzalez-alfonso17}--, and the ${\rm EW}^{~v<-200~{\rm km~s^{-1}}}_{\rm OH~119~\mu m}$ scaled by $L_{\rm FIR}^{1/2}$.  The best linear fit, forced to have zero intercept, yields

\begin{equation}\label{eq:EW}
\dot{M}_{\rm mol,out}~(M_{\odot}~{\rm yr}^{-1})=0.29\times\Bigg[\frac{{\rm EW}^{~v<-200~{\rm km~s^{-1}}}_{\rm OH~119~{\rm \mu m}}}{\rm km~s^{-1}}\times\Bigg(\frac{L_{\rm FIR}}{10^{10}~L_{\odot}}\Bigg)^{1/2}\Bigg],
\end{equation}

\noindent and is shown as a black line in Figure~\ref{fig:EW_dMdt}. On top the best-fit line we show the implied molecular mass outflow rate for IRAS~F13352+6402, IRAS~F20036-1547, and QSO~J131911+095051.

\end{appendix}

%-------------------------------------------------------------------

\bibliographystyle{aa}
\bibliography{references.bib}

\end{document}